\documentclass[preprint,fleqn,showpacs,showkeys, nodayofweek]{revtex4}

\usepackage{graphicx}
\usepackage{amssymb}
\usepackage{amsmath}
\usepackage{color}
\usepackage{bm}
\usepackage{datetime}
\begin{document}
\title{Late-time acceleration in the  coupled Cubic Galileon models
}

\author{Kyoung Yee Kim}
\email{kimky@inje.ac.kr}
\author{Hyung Won Lee}
\email{hwlee@inje.ac.kr}
\author{Yun Soo Myung}
\email{ysmyung@inje.ac.kr} \affiliation{Institute of Basic Science
and Department of Computer Simulation, Inje University, Gimhae
621-749, Korea}

\begin{abstract}
We investigate  the  linearly  and quadratically coupled cubic
Galileon models that include linear potentials.  These models  may
explain the late-time acceleration. In these cases, we need two
equations of state parameter named the native and effective
equations of state to test whether  the universe is accelerating or
not because there is coupling between the cold dark matter and
Galileon. It turns out that there is no transition from accelerating
phase to phantom phase in the future.
\end{abstract}



\maketitle

\section{Introduction}
\label{sec:introduction}

Observational data indicate that our universe undergoes an
accelerating phase since the recent past~\cite{Perlmutter:1999jt}.
The cosmological constant  could be considered  as a candidate for
the dark energy to explain the observational result in  the
$\Lambda$CDM model.  However, this model has two problems of the
fine tuning and the coincidence and thus, an alternative candidate
was needed. One promising candidate is a dynamical dark energy model
based on  scalar field theory  which is dubbed the
quintessence~\cite{Wetterich:1987fm,Ratra:1987rm,Caldwell:1998je}.
This model  has a canonical kinetic term and thus,  a  scalar is
minimally coupled to gravity.  We wish to point out that the
cosmological constant has a constant equation of state
$\omega=p/\rho=-1$, while the scalar field model has a
time-dependent equation of state with $-1 \le \omega \le 1$.

Recently, Planck observation~\cite{Ade:2013zuv} has shown four
combined data on equation of state:  i)
$\omega=-1.13^{+0.13}_{-0.25}$ (95$\%$; Panck+WMAP+BAO) which is in
good agreement with a cosmological constant,  ii) $\omega=-1.09\pm
0.17$ (95$\%$; Panck+WMAP+Union2.1) that  is more consistent with a
cosmological constant, iii) $\omega=-1.13^{+0.13}_{-0.14}$ (95$\%$;
Panck+WMAP+SNLS) which favors the phantom phase ($\omega<-1$) at the
$2\sigma$ level, and iv) $\omega=-1.24^{+0.18}_{-0.19}$ (95$\%$;
Panck+WMAP+$H_0$) which is in tension with a cosmological constant
at more than the $2\sigma$ level. The last two might draw the
universe into  a phantom phase  about at the 2$\sigma$ level.
  However, if one uses the BAO data in
addition to the CMB, there is  no strong evidence for the phantom
phase  that is  incompatible with a cosmological constant.

On the other hand, one  modified  gravity named the Galileon gravity
was considered another model of the dynamical dark
energy~\cite{Chow:2009fm}.    This model is also  described by  a
scalar field theory which contains terms of  nonlinear-derivative
self couplings.  Turning off gravity, the Galileon action  is
invariant under the Galilean transformation  of $\pi\to \pi+c+b_\mu
x^\mu$. The field equations contain at most second derivatives of
$\pi$, implying that it is surely  free from  the Ostrogradsky
ghosts. Turning on gravity, however,  breaks the symmetry.
Therefore, a covariant Galilean action has been
constructed~\cite{Deffayet:2009wt}, where the Galilean symmetry is
softly broken  but it preserves the shift symmetry of $\pi \to \pi
+c$.  Its cosmological implications   have been extensively
investigated to explain the late-time acceleration
in~\cite{Silva:2009km,DeFelice:2010pv,Gannouji:2010au,DeFelice:2011bh,Leon:2012mt,Adak:2013vwa}.

In this work, we wish to investigate  the late-time acceleration
 by using the linearly coupled cubic Galileon
model~\cite{Appleby:2011aa} together with a linear potential $V=c_1
\pi$.  This model is similar to the DGP model~\cite{Dvali:2000hr}.
We note that adding the potential breaks the shift symmetry, which
might make the nonlinear-derivative self coupling term  trivial.  In
this case, one  needs two equations of state parameter named the
native and effective equations of state to say whether the universe
is accelerating or not. This is  because there is coupling between
the cold dark matter and Galileon.   In the uncoupled case of
$c_0=0$, the model under consideration  corresponds to the cubic
Galileon gravity. This has been  studied in~\cite{Hossain:2012qm},
which shows that the data of SN+BAO+$H_0$ prefers the Galileon
gravity over the quintessence.  The authors in~\cite{Bartolo:2013ws}
have shown an appearance of the phantom phase, but it arose from
choosing negative $c_2$ and $c_3$.  More recently, it turned out
that the equation of state obtained from  the cubic Galileon model
is indistinguishable from $\omega=-1$ of the cosmological
constant~\cite{Bellini:2013hea}.   For $c_4\not=0$ and $c_5\not=0$,
a  phantom phase appeared for both $c_0=0$
case~\cite{Nesseris:2010pc} and $c_0\not=0$~\cite{Appleby:2011aa}.
For a quadratic coupling with $c_4\not=0$ and $c_5\not=0$, its
late-time evolution was investigated in~\cite{Jamil:2013yc} which
indicates a crossing of the phantom divide line.  However, we
mention that these phantom phases arose from  when one chooses
negative coefficients $c_i (i\ge 2)$.

Previous works on the Galilean models have shown phantom phases,
depending on the choice of coefficients. Hence, it is very curious
to check if a phantom phase  happens really in the Galilean models
because phantom behavior is quite interesting and it is possibly bad
for the future evolution of the universe. Hence, we hope to find a
phantom phase in the linearly and quadratically coupled cubic
Galileon model that include linear potential. However, we have found
no phantom phase when we did make a complete computation. We wish to
understand why there is no phantom by comparing it with the
Brans-Dicke theory with a linear potential.  If one suppresses a
nonlinear-derivative self coupling term of $(\nabla\pi)^2\square
\pi$ capturing   a decisive feature of the Galileon gravity, the
linearly coupled cubic Galileon model with the potential is similar
to the Brans-Dicke cosmology with the power-law potential
$\Phi^\alpha$ where one could observe a future crossing of the
phantom divide line only for $\alpha>1$~\cite{Lee:2010tm}.  Thus, it
seems   that  choosing the linear potential $V=c_1 \pi$ does not
provide the phantom phase because this potential breaks the shift
symmetry. Explicitly, the addition of the linear potential did not
make the nonlinear-derivative self coupling term of
$(\nabla\pi)^2\square
 \pi$ matter and thus,  it did not play a role in the late-time
 evolution.
  Very recently, a model of Slotheon gravity with $V=V_0e^{-\frac{\lambda}{M_{\rm pl}} \pi}$ and $c_0=c_2=c_3=c_4$
 has provided  the
cosmic acceleration but not the phantom phase in the late-time
evolution~\cite{Adak:2013vwa}.

Inspired by  the above motivation, we will focus on observing
whether the phantom phase appears in the future when we use the
linearly and quadratically coupled cubic Galileon models that
include linear potentials.

\section{Evolution equations}
\label{sec:equations} The covariant Galileon action can be written
as~\cite{Deffayet:2009wt}
\begin{equation}
\label{cov-galileon-action} S_{\rm cG} = \int d^4 x \sqrt{-g} \left
[ \frac{M_{\rm pl}^2R}{2} + \frac{1}{2}\Sigma_{i=1}^{5} c_i {\cal
L}_i + {\cal L}_{\rm m}-\frac{c_G}{M_{\rm pl}
M^3}T^{\mu\nu}\partial_\mu \pi\partial_\nu \pi-\frac{c_o}{M_{\rm
pl}} \pi T\right ],
\end{equation}
where $c_{1-5}$ are arbitrary dimensionless constants, and $M_{\rm
pl}$ is the reduced Planck mass, and $M^3=M_{\rm pl}H_0^2$ to make
the $c$'s dimensionless. ${\cal L}_{\rm m}$ denotes the Lagrangian
for the cold dark matter  and $T^{\mu\nu}$ (its trace $T$) represent
the energy-momentum tensor.  The Galileon action is usually
classified into three classes: the uncoupled Galileon  with
$c_0=c_G=0$; the linearly coupled Galileon with $c_0\not=0$ and
$c_G=0$; the derivative coupled Galileon with $c_0=0$ and
$c_G\not=0$.

 We are working in the Jordan frame where
the explicit coupling between $\pi$  and $T$ is removed when we use
a metric redefinition~\cite{Appleby:2011aa}. Among three classes of
the Galileon model, we focus on the linearly coupled cubic Galileon
(lcG) model as~\cite{Hossain:2012qm}
\begin{equation}
\label{galileon-action} S_{\rm lcG} = \int d^4 x \sqrt{-g} \left [
\left( 1-2c_0 \frac{\pi}{M_{\rm pl}} \right ) \frac{M_{\rm
pl}^2R}{2} - \frac{c_2}{2}(\nabla \pi)^2-\frac{c_3}{M^3}(\nabla
\pi)^2\square \pi - V(\pi) + {\cal L}_{\rm m} \right ],
\end{equation}
where  the linear potential of $V(\pi)=c_1\pi$ is introduced for our
late-time evolution. Here we demand  that $c_i$ $(i\ge 2)$ are
positive to avoid the phantom scalar field.  The case of $c_0=0$
corresponds to the cubic Galileon gravity~\cite{Bellini:2013hea}.
This model with $V(\pi)=0$ was extensively studied
in~\cite{Appleby:2011aa} and the uncoupled case of $c_0=0$, $c_2=1$
and $c_3/M^3=\alpha/2M^3_{\rm pl}$ was investigated
in~\cite{Hossain:2012qm}. In addition to the choice of $c_2=1$, we
use $\alpha$ instead of $c_3$ which shows a decisive feature of the
Galileon gravity when we compare it with the quintessence and the
Brans-Dicke cosmology.

The Einstein equation is derived from (\ref{galileon-action}) as
\begin{equation} \label{ein-eq0}
M^{2}_{\rm pl}\Big(1-\frac{2c_0}{M_{\rm pl}}\pi\Big)
G_{\mu\nu}=\Big[-2M_{\rm pl}c_0
(\nabla_\mu\nabla_\nu-g_{\mu\nu}\nabla^2)\pi
+T_{\mu\nu}+T^{\alpha}_{\mu\nu}+T^{V}_{\mu\nu}+T^{ m}_{\mu\nu}\Big],
\end{equation}
where $G_{\mu\nu}=R_{\mu\nu}- Rg_{\mu\nu}/2$ is the Einstein tensor
and the energy-momentum tensors  are defined to be
~\cite{Appleby:2011aa}
\begin{eqnarray}
T_{\mu\nu}&=& \Big[\nabla_\mu \pi \nabla_\nu \pi
-\frac{1}{2} g_{\mu\nu} (\nabla \pi)^2\Big], \\
T^{\alpha}_{\mu\nu}&=& \frac{\alpha}{M_{\rm pl}^3}\Big[\nabla_\mu
\pi \nabla_\nu \pi \square \pi +\frac{g_{\mu\nu}}{2} \nabla_\alpha
\pi \nabla^\alpha (\nabla
\pi)^2-\nabla_{(\mu}\pi\nabla_{\nu)}(\nabla \pi)^2\Big], \\
T^{V}_{\mu\nu}&=&\frac{c_1}{2} \pi g_{\mu\nu}.
\end{eqnarray}
Here the energy-momentum tensor $T^{ m}_{\mu\nu}$ for a pressureless
matter of  ${\cal L}_{\rm m} $ is given by
\begin{equation}
T^{m}_{\mu\nu}=\rho_{\rm m} u_\mu u_\nu \end{equation} with $u_\mu$
the four velocity.

The Galileon equation takes the form
\begin{equation} \label{gal-eq}
\square \pi +\frac{\alpha}{M_{\rm pl}^3} \Big[(\square
\pi)^2-(\nabla_\mu\nabla_\nu \pi)^2-R^{\mu\nu}\nabla_\mu\pi
\nabla_\nu\pi\Big] -V_{,\pi}-c_0 M_{\rm pl} R=0,
\end{equation}
where  $V_{,\pi}$ denotes the derivative with respect to $\pi$. To
study its cosmological implications, it would be better to convert
Eq. (\ref{ein-eq0})  into a standard  form of the Einstein equation
\begin{equation} \label{ein-eq}
G_{\mu\nu}=M^{-2}_{\rm pl}\Bigg(T^{\rm \pi}_{\mu\nu}+\frac{T^{
m}_{\mu\nu}}{1- \frac{2c_0}{M_{\rm pl}}\pi}\Bigg),
\end{equation}
where  $T^{\pi}_{\mu\nu}$ denotes the energy-momentum tensor from
all $\pi$'s contributions$\times (1-2c_0\pi/M_{\rm pl})^{-1}$.

Importantly, the Bianchi identity obtained by acting $\nabla^\mu$ on
Eq. (\ref{ein-eq}) leads to the total conservation-law
as~\cite{Lee:2010tm}
\begin{equation} \label{con-law}
\nabla^\mu T^{\rm \pi}_{\mu\nu}+  T^{\rm
m}_{\mu\nu}\nabla^\mu\Big(\frac{1}{1-\frac{2c_0}{M_{\rm
pl}}\pi}\Big)=0
\end{equation}
when one uses the conservation-law for the cold dark matter
\begin{equation}
\nabla^\mu T^{ m}_{\mu\nu}=0.
\end{equation}

At this stage, we introduce the flat Friedmann-Robertson-Walker
(FRW) metric as
\begin{equation}
\label{metric} ds^2 = -dt^2 + a^2(t) \left ( dr^2 + r^2 d\theta^2
+ r^2 \sin^2\theta d\varphi^2 \right ),
\end{equation}
where $a(t)$ is the scale factor.
 Then, we write the Einstein equation (\ref{ein-eq0}) and the Galileon
 equation (\ref{gal-eq}) as
\begin{eqnarray}
\label{hubble1} &&\hspace{-5mm}3 \left ( 1 - \frac{2c_0}{M_{\rm pl}}\pi\right )
M_{\rm pl}^2H^2 = \rho_{\rm m} +
6 c_0 M_{\rm pl} H {\dot \pi} +
\frac{\dot \pi^2}{2} \left ( 1-6 \frac{\alpha}{M_{\rm pl}^3} H \dot \pi \right ) + V(\pi),\\
\label{hubble2} &&\hspace{-5mm}\left ( 1 - \frac{2c_0}{M_{\rm pl}}\pi\right )
M_{\rm pl^2} \left ( 2 \dot H + 3 H^2 \right ) =
 2 c_0 M_{\rm pl} \left ( {\ddot \pi} + 2 H {\dot \pi} \right )
-\frac{\dot \pi^2}{2} \left ( 1+2 \frac{\alpha}{M_{\rm pl}^3} \ddot \pi \right ) + V(\pi),\\
\label{hubble3} &&\hspace{-5mm} {\ddot \pi} +3 H {\dot \pi}  -
\frac{3\alpha}{M_{\rm pl}^3} {\dot \pi} \left ( 3 H^2 {\dot \pi} +
{\dot H} {\dot \pi} + 2H {\ddot \pi} \right ) + V_{,\pi} + 6 M_{\rm
pl} c_0 \left ( 2H^2 + {\dot H} \right )=0.
\end{eqnarray}
In the case of $c_0=0$, the above all equations reduce  to Eqs.
(2)-(4) of Ref.~\cite{Hossain:2012qm}.  We can rewrite Eqs.
(\ref{hubble1}) and (\ref{hubble2}) as the standard forms of the
Friedmann equations
\begin{eqnarray}
\label{hubble1m} &&3  M_{\rm pl}^2H^2 = \frac{\rho_{\rm m}}{\left (
1 - \frac{2 c_0}{M_{\rm pl}}\pi\right )} + \frac{\left \{
6 c_0 M_{\rm pl} H {\dot \pi} +
\frac{\dot \pi^2}{2} \left ( 1-6 \frac{\alpha}{M_{\rm pl}^3} H \dot \pi \right )
+ V(\pi) \right \}}{\left ( 1 -
\frac{2c_0}{M_{\rm pl}}\pi \right )} ,\\
\label{hubble2m} &&M_{\rm pl}^2 \left ( 2 \dot H + 3 H^2 \right ) =
- \frac{\left
\{ - 2 c_0 M_{\rm pl} \left ( {\ddot \pi} + 2 H {\dot \pi} \right ) + \frac{\dot \pi^2}{2} \left ( 1+2 \frac{\alpha}{M_{\rm pl}^3}
\ddot \pi \right ) - V(\pi) \right \}}{\left ( 1 - \frac{2c_0}{M_{\rm pl}}\pi \right ) } ,
\end{eqnarray}
which can also found from (\ref{ein-eq}) directly.  Since Eqs.
(\ref{hubble1m}) and (\ref{hubble2m}) are the first and second
Friedmann equations, respectively,  we can read off  the energy
density and pressure for the Galileon   as
\begin{eqnarray}
\label{rho_pi} \rho_{\pi} &=& \frac{1}{\left ( 1 -
\frac{2c_0}{M_{\rm pl}}\pi \right )} \left \{
6 c_0 M_{\rm pl} H {\dot \pi} +
\frac{\dot \pi^2}{2} \left ( 1-6 \frac{\alpha}{M_{\rm pl}^3} H \dot \pi \right ) + V(\pi) \right \}, \\
\label{p_pi} p_\pi &=& \frac{1}{\left ( 1 -\frac{2 c_0 }{M_{\rm
pl}}\pi \right ) } \left \{
- 2 c_0 M_{\rm pl} \left ( {\ddot \pi} + 2 H {\dot \pi} \right )
+\frac{\dot \pi^2}{2} \left ( 1+2
\frac{\alpha}{M_{\rm pl}^3} \ddot \pi \right ) - V(\pi) \right \}.
\end{eqnarray}

Now we express  the total conservation-law (\ref{con-law}) in terms
of density and pressure for the Galileon and  matter density as
\begin{equation}
\label{conserv-eq} {\dot \rho}_\pi + 3 H \left ( \rho_\pi + p_\pi
\right ) = - \frac{2 c_0 \frac{\dot \pi}{M_{\rm pl}}\rho_{\rm m}}
{\left ( 1 - \frac{2 c_0 }{M_{\rm pl}}\pi \right)^2},
\end{equation}
which is very similar to the Brans-Dicke
cosmology~\cite{Lee:2010tm}. We observe that the left-hand side of
Eq. (\ref{conserv-eq})  implies an energy (matter) transfer between
Galileon and cold dark matter.

\section{Background Evolution}

To solve three coupled
 equations (\ref{hubble1})-(\ref{hubble3}), we introduce  the new
 variables
\begin{eqnarray}
&& x = \frac{{\dot \pi}}{\sqrt{6} H M_{\rm pl}}, ~~ y =
\frac{\sqrt{V}}{\sqrt{3} H M_{\rm pl}},
~~ \epsilon = -\frac{6 H {\dot \pi}}{M_{\rm pl}^3}, \\
&& \lambda = - M_{\rm pl} \frac{V_{,\pi}}{V}, ~~ z = - \frac{2
\pi}{M_{\rm pl}}, ~~ \eta = \ln \Big[\frac{a}{a_0}\Big],
\end{eqnarray}
where $\eta$ is introduced instead of the scale factor $a$.
Equations (\ref{hubble1})-(\ref{hubble3})
 are transformed into the first-order coupled equations
\begin{eqnarray}
x' &=& \frac{-1}{1+\alpha \epsilon} \left \{ 3x + \frac{3}{2}
\alpha \epsilon x - \frac{\sqrt{6}}{2}\lambda y^2 + 2 \sqrt{6} c_0
+ \frac{H'}{H} \left ( x + \frac{3}{2} \alpha \epsilon x + \sqrt{6} c_0 \right )\right \}, \\
y' &=& -y \left ( \frac{H'}{H} + \frac{\sqrt{6}}{2} \lambda x \right ), \\
\epsilon' &=& \epsilon \left ( 2 \frac{H'}{H} + \frac{x'}{x}\right ), \\
\lambda ' &=& \sqrt{6} \lambda^2 x \left ( 1 - \Gamma \right ), \\
z' &=& -2 \sqrt{6} x,\\
\frac{H'}{H} &=& \frac{A} {B},
\end{eqnarray}
where $\Gamma = \frac{V V_{, \pi\pi}}{V_{,\pi}^2}$ and  prime ($'$)
denotes the differentiation  with respect to $\eta$. Here,  $A$ and
$B$ are given as
\begin{eqnarray}
A &=& -3(2+4 \alpha \epsilon + \alpha^2
\epsilon^2)x^2 +
\sqrt{6} \left ( \lambda y^2 \alpha \epsilon - 2 c_0 \alpha \epsilon - 4 c_0 \right ) x
\nonumber \\
&&+
6 ( 1+\alpha \epsilon) (y^2 -1 -c_0 z) + 12 c_0 \lambda y^2 - 48 c_0^2,
\\
B &=&
\alpha^2 \epsilon^2 x^2 +
4 \sqrt{6} c_0 \alpha \epsilon  x +
4 \left ( 1+ c_0 z \right ) \left ( 1+ \alpha \epsilon \right ) +
24 c_0^2.
\end{eqnarray}

Importantly, the total conservation-law  (\ref{conserv-eq}) can be
rewritten as the new variables as
\begin{equation}
\label{conserv-eqm} {\dot \rho}_\pi + 3 H \left ( 1+
\omega_{\pi}^{\rm nat} \right ) \rho_\pi = H \left ( -2 \sqrt{6}
\frac{c_0 x \rho_{\rm m}}{\left (1 + c_0 z \right )^2}\right ),
\end{equation}
where the native equation of state is defined to be
\begin{equation}
\omega_{\pi}^{\rm nat} = \frac{p_{\pi}}{\rho_{\pi}} = \frac{
-\frac{2 \sqrt{6}}{3} c_0 x \left ( 2 + \frac{H'}{H} + \frac{x'}{x} \right )
+ x^2 \left (1  -\frac{1}{3} \alpha \epsilon \frac{H'}{H}
      -\frac{1}{3} \alpha \epsilon \frac{x'}{x}\right ) - y^2}
{2 \sqrt{6} c_0 x + x^2 (1 + \alpha \epsilon )  + y^2}.
\end{equation}
One may  interpret Eq. (\ref{conserv-eqm}) as
\begin{equation}
\label{conserv-eq0} {\dot \rho}_\pi + 3 H \left ( 1+
\omega_{\pi}^{\rm nat}
   +\frac{2 \sqrt{6}}{3} \frac{c_0 x }{\left (1 + c_0 z \right )^2}
      \frac{\rho_{\rm m}}{\rho_\pi}
\right ) \rho_\pi = 0.
\end{equation}
From this equation, one can obtain
 the effective equations of state as
\begin{equation}
\omega_{\pi}^{\rm eff} = \omega_{\pi}^{\rm nat} +
\frac{2\sqrt{6}}{3} \frac{c_0 x}{1+ c_0 z}
    \frac{1-\Omega_{\rm m}}{\Omega_{\pi}} ,
\end{equation}
where the two density parameter $\Omega_{\pi}$ are  given by
\begin{equation}
\label{Omgea_pi} \Omega_{\pi} = \frac{\rho_{\pi}}{\rho_c}= \frac{ 2
\sqrt{6} c_0 x + x^2 \left ( 1+ \alpha \epsilon\right ) + y^2}{1+c_0
z},~~ \Omega_{\rm m}=  1 - \Omega_{\pi}= \frac{\rho_{\rm m}}{\rho_c
\left ( 1+ c_0 z\right )}.
\end{equation}
Here $\rho_c = 3 M_{\rm pl}^2 H^2$ is the critical energy density
and we  used the relation
\begin{equation}
\frac{1}{1+c_0 z} \frac{\rho_{\rm m}}{\rho_\pi} = \frac{\Omega_{\rm
m}}{\Omega_\pi}.
\end{equation}
In case of $c_0=0$, we have $\omega_{\pi}^{\rm
nat}=\omega_{\pi}^{\rm eff}$, which implies that $\omega_{\pi}^{\rm
eff}$ is not necessary to describe  the uncoupled Galileon model.
Since  $\omega^{\rm eff}_{\pi}$ and $\Omega_{\rm m}$ blow up at
$z=-1/c_0$ for $c_0<0$, we  require   $c_0>0$.

\begin{figure}
\includegraphics[width=0.6\textwidth]{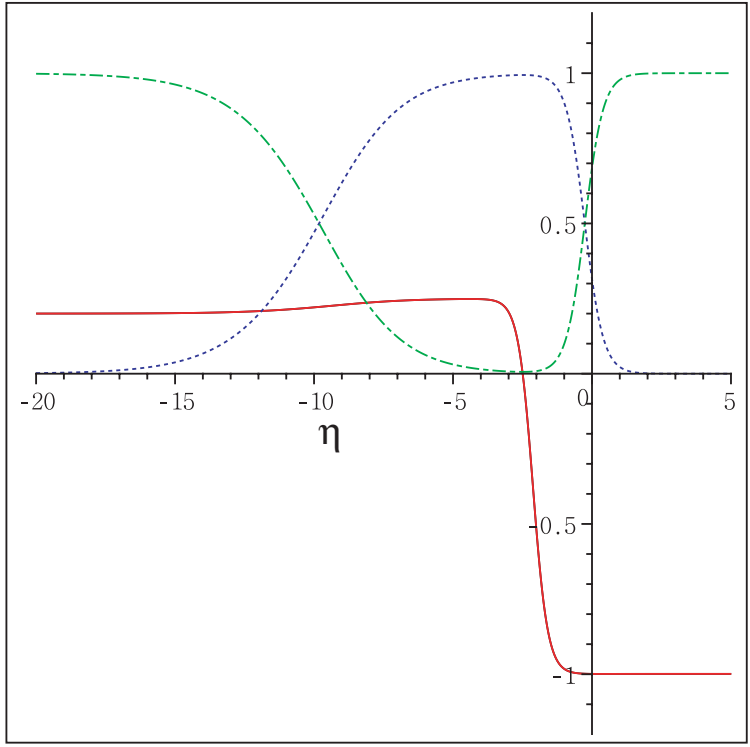}     % includes figure foo.eps
\caption{Evolutions for the uncoupled Galileon with $c_0 = 0$.
$\eta=\ln[a/a_0]<0(\eta>0)$ denote the past (future) and $\eta=0$
represents the present time with $a=a_0$. These include density
parameter $\Omega_{\rm m}$ of cold dark matter (blue-dotted) and
density parameter $\Omega_{\rm \pi}$ of Galileon
(green-dotted-dashed). Red-solid curve denotes the equation of state
$\omega^{\rm nat}_{\rm \pi}$ for Galileon. We impose $\Omega_{\rm
m}^0=0.315$ at $\eta=0$ as an initial condition and $\alpha = 1.0$.
The other initial conditions are given by $\epsilon_0 = 5.0$, $x_0 =
0.01$, $y_0 = 0.8272847152$, $\lambda_0 = 0.1$ and $z_0 = 1.0$.}
\label{fig1}
\end{figure}

\begin{figure}
\includegraphics[width=0.6\textwidth]{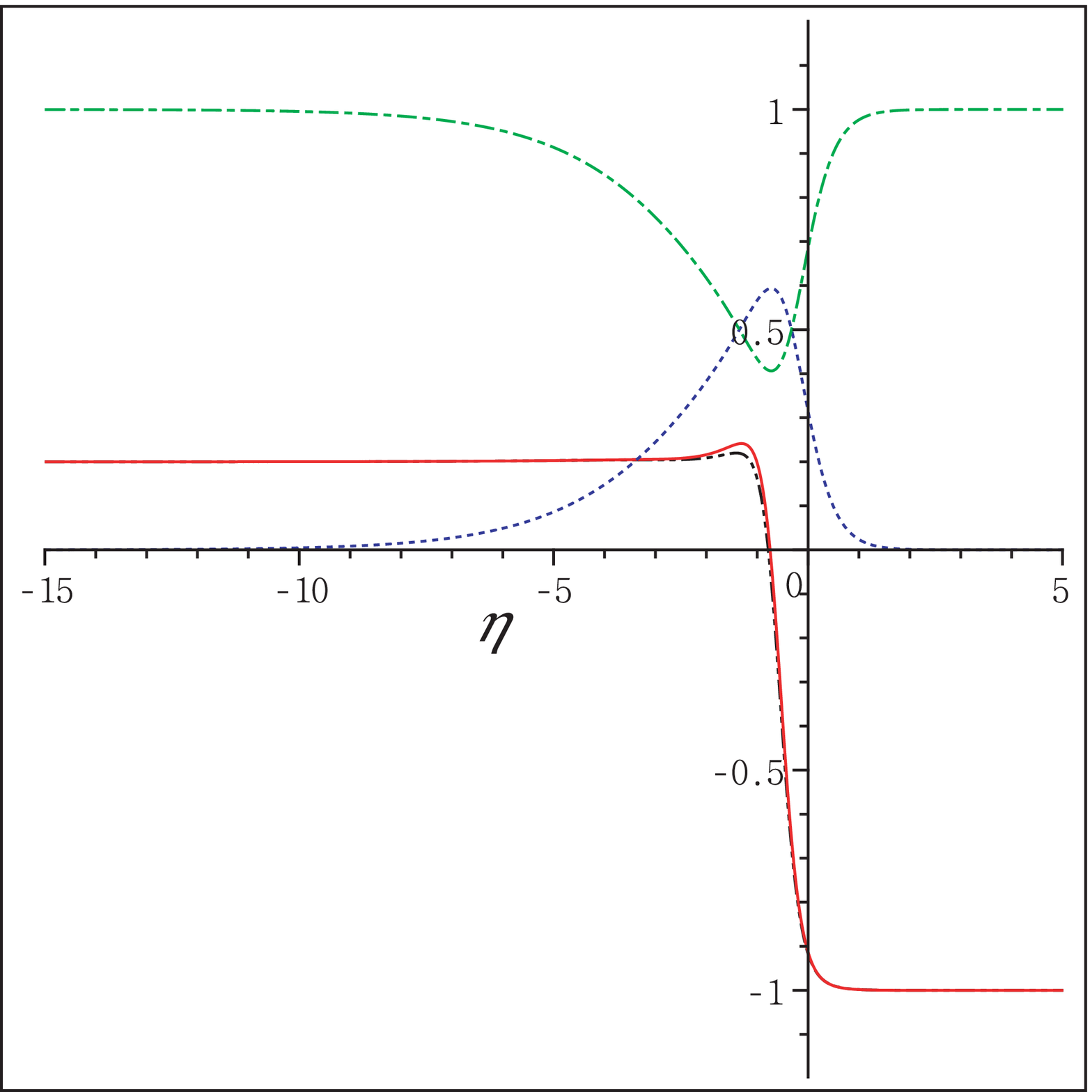}     % includes figure foo.eps
\caption{Evolutions for the linearly coupled Galileon with $c_0 =
0.1$.  These include density parameter $\Omega_{\rm m}$ of cold dark
matter (blue-dotted) and density parameter $\Omega_{\rm \pi}$ of
Galileon (green-dotted-dashed) as functions of $\eta=\ln[a/a_0]$.
Black-solid (red-dotted-dashed) curves denote the equation of state
$\omega^{\rm nat}_{\rm \pi}$ ($\omega^{\rm eff}_{\rm \pi}$) for
Galileon. We choose $\Omega^0_{\rm m}=0.315$ at $\eta=0$ and $\alpha
= 1.0$. The initial conditions are $\epsilon_0 = 0.1, x_0 = 0.1$,
$y_0 = 1.144556772$, $\lambda_0 = 0.1$ and $z_0 =10.0$.}
\label{fig2}
\end{figure}

We rewrite  the first Friedmann equation (\ref{hubble1}) in terms of
the new variables as
\begin{equation}
\label{Omega_m} \Omega_{\rm m} = 1 - \frac{2 \sqrt{6} c_0 x + x^2
\left ( 1 + \alpha \epsilon \right ) + y^2}{1+c_0 z},
\end{equation}
which was  used  to find the initial values of  the evolving
variables. \label{sec:backhround} According to the  Planck
mission~\cite{Ade:2013zuv}, the current dark matter content is
$\Omega^0_{\rm m} = 0.315$. We take  this value as an initial
condition for  the numerical evolution.  After solving the
first-order coupled equations, one finds  the background evolution
for the uncoupled case ($c_0 = 0$). As is depicted  in  Fig.
\ref{fig1}, we observe the accelerating universe of  $\omega_{\rm
\pi}^{\rm nat} \ge -1$ in the future. Furthermore, Fig. \ref{fig2}
shows a typical background evolution  for the linearly coupled case
($c_0 \ne 0$). In this case, we find the accelerating universe which
is shown  by $\omega_{\rm \pi}^{\rm nat} \ge -1$ as well as
$\omega_{\rm \pi}^{\rm eff} \ge -1$.  Definitely, there is  no
signal to give a  phantom phase with $\omega_{\rm \pi}^{\rm nat}<-1$
in the future.

\section{Quadratic Coupling}
\label{qudratic} 
\begin{figure}
\includegraphics[width=0.6\textwidth]{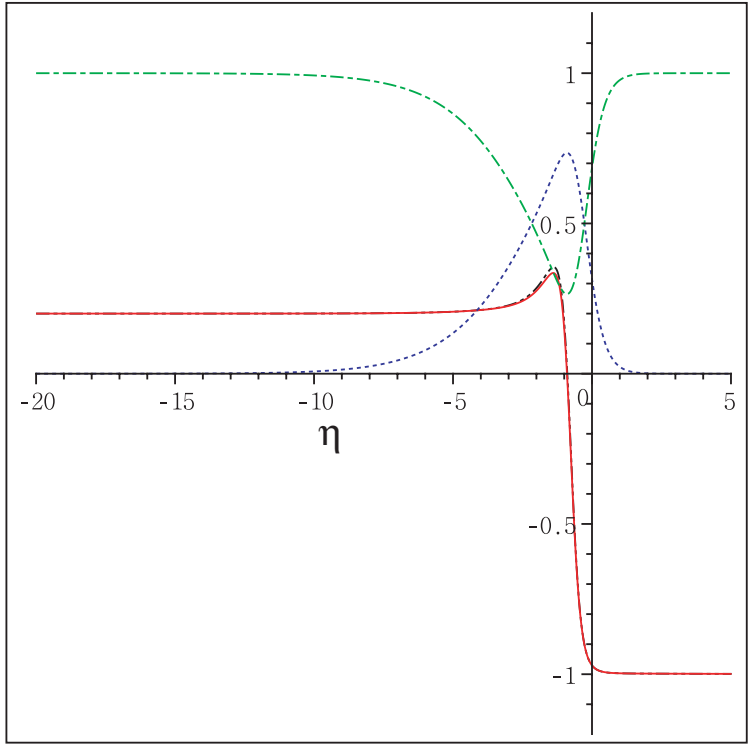}     % includes figure foo.eps
\caption{Evolutions for the quadratically coupled Galileon with $c_0
= 0.01$.  These include density parameter $\Omega_{\rm m}$ of cold
dark matter (blue-dotted) and density parameter $\Omega_{\rm \pi}$
of Galileon (green-dotted-dashed) as functions of $\eta=\ln[a/a_0]$.
Black-solid (red-dotted-dashed) curves denote the equation of state
$\omega^{\rm nat}_{\rm qc\pi}$ ($\omega^{\rm eff}_{\rm qc\pi}$) for
Galileon. We choose $\Omega_{\rm m}^0=0.315$ at $\eta=0$ and $\alpha
= 1.0$. The other initial conditions are $\epsilon_0 = 0.1, x_0 =
0.1$, $y_0 = 0.8209750301$, $\lambda_0 = 0.1$ and $z_0 =0.0$.}
\label{fig3}
\end{figure}

Recently, there was a work for the quadratic
coupling which shows a crossing of the phantom
divide~\cite{Jamil:2013yc}. To see whether this  happens really or
not, we consider the  action without ${\cal L}_4$ and ${\cal L}_5$
for simplicity
\begin{equation}
\label{galileon2-action} S_{\rm qcG} = \int d^4 x \sqrt{-g} \left [
\left( 1-4c_0 \frac{\pi^2}{M_{\rm pl}^2} \right ) \frac{M_{\rm
pl}^2R}{2} - \frac{c_2}{2}(\nabla \pi)^2-\frac{c_3}{M^3}(\nabla
\pi)^2\square \pi - V(\pi) + {\cal L}_{\rm m} \right ].
\end{equation}
Following the computation steps of the previous section, two
equations that are  different from the linearly coupled case are
\begin{eqnarray}
x' &=& \frac{-1}{1+\alpha \epsilon} \left \{ 3x + \frac{3}{2}
\alpha \epsilon x - \frac{\sqrt{6}}{2}\lambda y^2 - 4 \sqrt{6} c_0 z
+ \frac{H'}{H} \left ( x + \frac{3}{2} \alpha \epsilon x - 2 \sqrt{6} c_0 z \right )\right \}, \\
\frac{H'}{H} &=& \frac{A_{\rm qc}} {B_{\rm qc}},
\end{eqnarray}
where $A_{\rm qc}$ and $B_{\rm qc}$ are given as
\begin{eqnarray}
A_{\rm qc} &=& -3(2+4 \alpha \epsilon + \alpha^2
\epsilon^2 - 32 c_0 \alpha \epsilon - 32 c_0 )x^2 +
\sqrt{6} \left ( \lambda y^2 \alpha \epsilon + 4 c_0 z \alpha \epsilon + 8 c_0 z \right ) x
\nonumber \\
&&+
6 ( 1+\alpha \epsilon) (y^2 -1 + c_0 z^2) - 24 c_0 z \lambda y^2 - 192 c_0^2 z^2,
\\
B_{\rm qc} &=&
\alpha^2 \epsilon^2 x^2 -
8 \sqrt{6} c_0 z \alpha \epsilon x +
4 \left ( 1- c_0 z^2 \right ) \left ( 1+ \alpha \epsilon \right ) +
96 c_0^2 z^2.
\end{eqnarray}
Also, the native and effective equations of state are given as
\begin{equation}
\omega_{{\rm qc}\pi}^{\rm nat} = \frac{p_{\rm qc\pi}}{\rho_{\rm
qc\pi}} = \frac{ \frac{4 \sqrt{6}}{3} c_0 z x \left ( 2 +
\frac{H'}{H} + \frac{x'}{x} \right ) - 16 c_0 x^2 + x^2 \left (1
-\frac{1}{3} \alpha \epsilon \frac{H'}{H}
      -\frac{1}{3} \alpha \epsilon \frac{x'}{x}\right ) - y^2}
{-4 \sqrt{6} c_0 z x + x^2 (1 + \alpha \epsilon )  + y^2},
\end{equation}
\begin{equation}
\omega_{{\rm qc}\pi}^{\rm eff} = \omega_{{\rm qc}\pi}^{\rm nat}
-\frac{4 \sqrt{6}}{3} \frac{c_0 z x}{1-c_0 z^2} \frac{1-\Omega_{\pi}}{\Omega_{\pi}} .
\end{equation}
We note that the first Friedmann equation for the initial condition
is slightly modified to be
\begin{equation}
\label{Omega_m2} \Omega_{\rm m} = 1 - \frac{-4 \sqrt{6} c_0 z x + x^2 \left ( 1 + \alpha
\epsilon \right ) + y^2}{1- c_0 z^2}.
\end{equation}
We observe from  Fig. \ref{fig3} that there is no crossing of the
phantom divide in the future. There is no essential difference
between the linearly coupled and quadratically coupled Galileon
models.

\section{Discussions}
\label{sec:discussions}

 First of all, we observe that  the native
and effective equations of state
 do not cross the phantom divide of  $\omega=-1$ in  the linearly (quadratically) coupled cubic Galileon models.
 This means that there is no essential difference between the Brans-Dicke cosmology and
 the cubic Galileon models. In other words,  the term of  $(\nabla\pi)^2\square \pi$
 showing a feature of the Galileon gravity did not contribute
 significantly to  deriving the late-time acceleration.
 To explain it, let us compare our result with
 Ref.~\cite{Appleby:2011aa} where acceleration was found, even though the potential $V=c_1\pi$ was not
 introduced. The reason seems to be clear because   this potential breaks the shift
 symmetry of $\pi \to \pi+c$. The addition of  the potential  did not make the nonlinear-derivative self coupling  term of $(\nabla\pi)^2\square
 \pi$ matter and thus,  it did not play a role in the late-time
 evolution.

  At this stage, one may ask  which one is an observable
quantity between $\omega_{\rm \pi}^{\rm eff}$ and $\omega_{\rm
\pi}^{\rm nat}$ in the Jordan frame~\cite{Lee:2010tm,Lee:2011tb}. It
is well known that the Jordan frame is a physical frame  because of
a minimal coupling to matter. However, this frame gives rises to the
non-conservation of continuity equation (\ref{conserv-eq}) which
shows that $\rho_{\rm m}$ plays the role of a source to generate a
new dark fluid.  Even though $\omega_{\rm \pi}^{\rm nat}$ indicates
a genuine equation of state for a Galileon-fluid, it cannot satisfy
the continuity equation.  On the other hand, although $\omega_{\rm
\pi}^{\rm eff}$ is not a genuine equation of state for a
Galileon-fluid (because it contains cold dark matter), it satisfies
the continuity equation.  This implies that each of them is not a
perfect observable for the linearly (quadratically)  coupled
Galileon models.  Therefore, we have to use both $\omega_{\rm
\pi}^{\rm nat}$ and $\omega_{\rm \pi}^{\rm eff}$ to show  the
presence   of a crossing of the phantom divide.  If the phantom
phase is observed  from both, one believes that it really happens in
the evolution of the linearly (quadratically) coupled cubic Galileon
gravity models.  Otherwise, one is hard to confirm the appearance of
the phantom phase in the Jordan frame. Surely, our analysis shows
the disappearance of any phantom phase in the coupled cubic Galileon
models.

\begin{figure}
\includegraphics[width=0.6\textwidth]{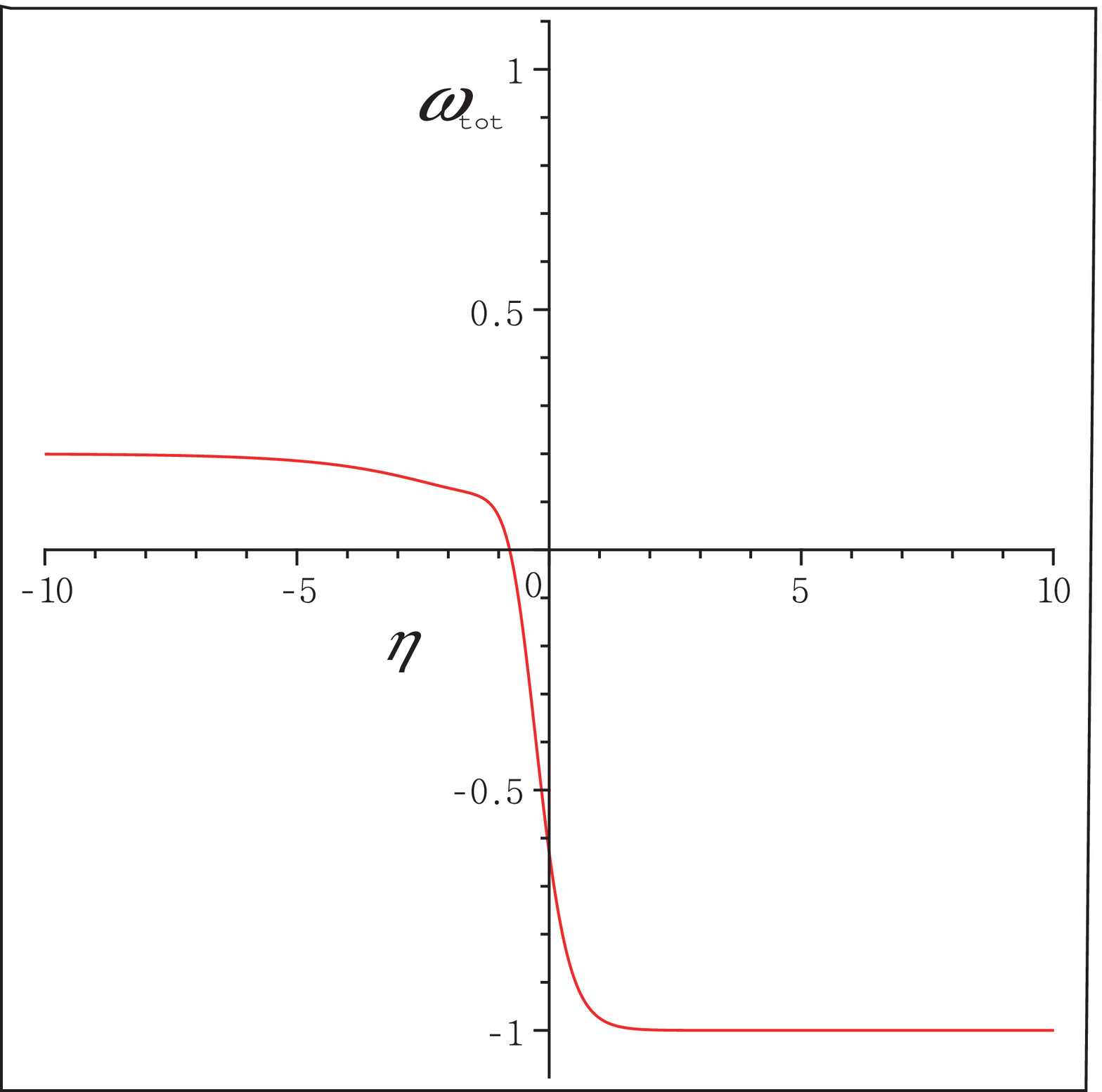}     % includes figure foo.eps
\caption{Evolution for $\omega_{\rm tot}$ (\ref{w_tot}) as a
function of $\eta=\ln[a/a_0]$ for the linearly coupled cubic
Galileon model. All other conditions are the same as the Fig. 2
caption does show.} \label{fig4}
\end{figure}

\begin{figure}
\includegraphics[width=0.6\textwidth]{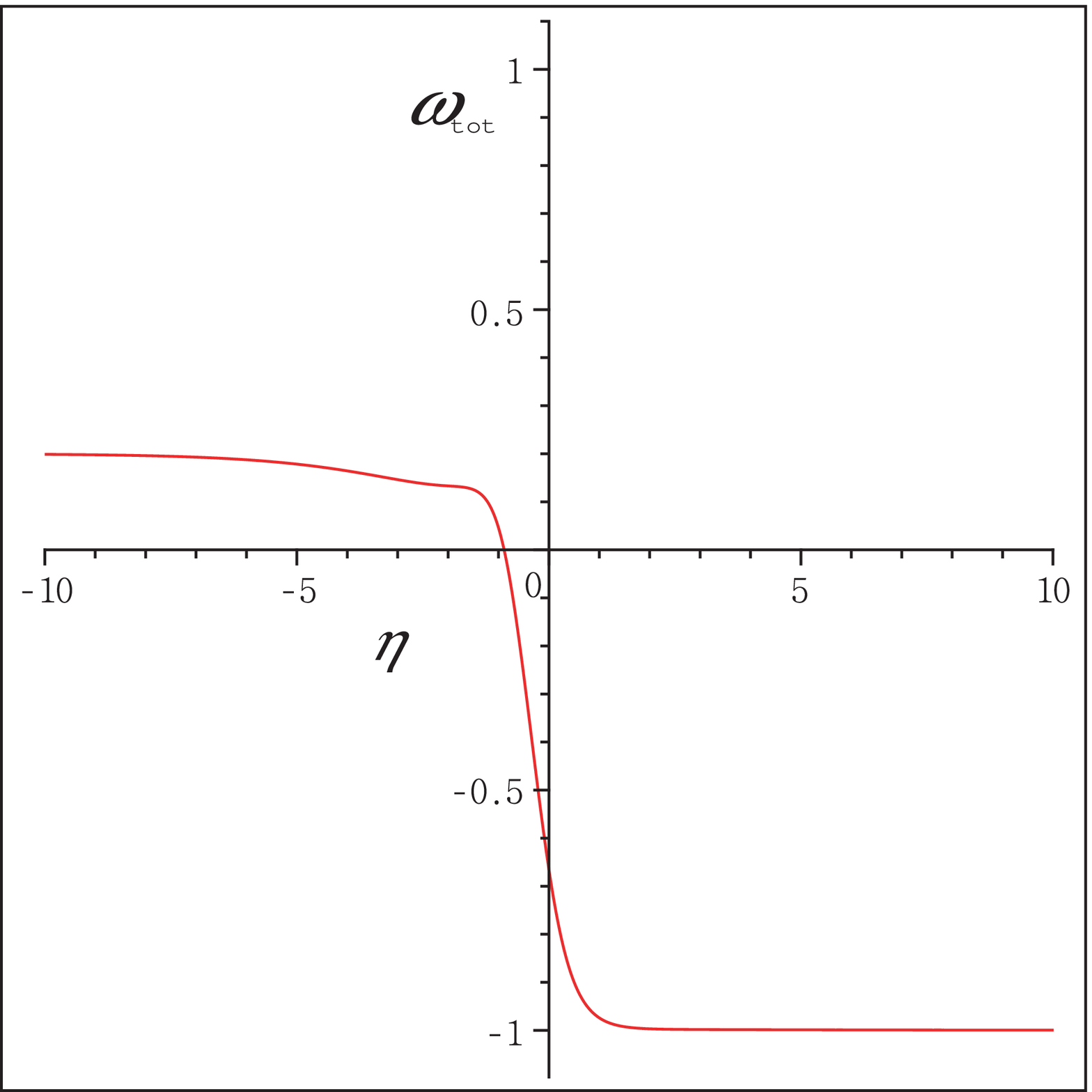}     % includes figure foo.eps
\caption{Evolution for $\omega_{\rm tot}$ (\ref{w_tot}) as a
function of $\eta=\ln[a/a_0]$ for the quadratically coupled cubic
Galileon model. All other conditions are the same as the Fig. 3
caption does show.} \label{fig5}
\end{figure}

Finally, we mention that   there is  the equation
~\cite{Chow:2009fm,Leon:2012mt,Lee:2010tm}
\begin{equation}
\frac{\dot{H}}{H^2}=-\frac{3}{2} (1+\omega_{\rm tot}).
\end{equation}
which defines  the total equation of state
\begin{equation}
\omega_{\rm
tot}=-1-\frac{2\dot{H}}{3H^2}=-1-\frac{2H'}{3H}=\frac{p_\pi}{\rho_{\rm
m}+\rho_\pi}.
\label{w_tot}
\end{equation}
The evolution of $\omega_{\rm tot}$ is  given in Figs. \ref{fig4}
and \ref{fig5} for linearly coupled and quadratically coupled cases,
respectively.  These  also  show  that there is no phantom phase in
the future.


%\appendix

%\section{}
%Let us go then, you and I\ldots

\acknowledgments This research was supported by Basic Science
Research Program through the National Research Foundation of Korea
(NRF) funded by the Ministry of Education, Science and Technology
(NO. 2012R1A1A3012776).


\begin{thebibliography}{99}

%\cite{Perlmutter:1999jt}
\bibitem{Perlmutter:1999jt}
  S.~Perlmutter, M.~S.~Turner and M.~J.~White,
  %``Constraining dark energy with SNe Ia and large scale structure,''
   Phys.\ Rev.\ Lett.\  {\bf 83}, 670 (1999)  [astro-ph/9901052].
    %%CITATION = ASTRO-PH/9901052;%%  %322 citations counted in INSPIRE as of 11 Jul 2013
%\cite{Wetterich:1987fm}
\bibitem{Wetterich:1987fm}
  C.~Wetterich,
  %``Cosmology and the Fate of Dilatation Symmetry,''
   Nucl.\ Phys.\ B {\bf 302}, 668 (1988).
   %%CITATION = NUPHA,B302,668;%%  %1463 citations counted in INSPIRE as of 11 Jul 2013

%\cite{Ratra:1987rm}
\bibitem{Ratra:1987rm}
  B.~Ratra and P.~J.~E.~Peebles,
  %``Cosmological Consequences of a Rolling Homogeneous Scalar Field,''
  Phys.\ Rev.\ D {\bf 37}, 3406 (1988).
   %%CITATION = PHRVA,D37,3406;%%  %2379 citations counted in INSPIRE as of 11 Jul 2013

%\cite{Caldwell:1998je}
\bibitem{Caldwell:1998je}
  R.~R.~Caldwell, R.~Dave and P.~J.~Steinhardt,
  %``Quintessential cosmology: Novel models of cosmological structure formation,''
  Astrophys.\ Space Sci.\  {\bf 261}, 303 (1998).
  %%CITATION = APSSB,261,303;%%  %35 citations counted in INSPIRE as of 11 Jul 2013

  %\cite{Ade:2013zuv}
\bibitem{Ade:2013zuv}
  P.~A.~R.~Ade {\it et al.}  [Planck Collaboration],
  %``Planck 2013 results. XVI. Cosmological parameters,''
  arXiv:1303.5076 [astro-ph.CO].
  %%CITATION = ARXIV:1303.5076;%%  %353 citations counted in INSPIRE as of 11 Jul 2013

%\cite{Chow:2009fm}
\bibitem{Chow:2009fm}
  N.~Chow and J.~Khoury,
  %``Galileon Cosmology,''
  Phys.\ Rev.\ D {\bf 80}, 024037 (2009)  [arXiv:0905.1325 [hep-th]].
  %%CITATION = ARXIV:0905.1325;%%  %111 citations counted in INSPIRE as of 12 Jul 2013



%\cite{Deffayet:2009wt}
\bibitem{Deffayet:2009wt}
  C.~Deffayet, G.~Esposito-Farese and A.~Vikman,
  %``Covariant Galileon,''
  Phys.\ Rev.\ D {\bf 79}, 084003 (2009)  [arXiv:0901.1314 [hep-th]].
   %%CITATION = ARXIV:0901.1314;%%  %207 citations counted in INSPIRE as of 11 Jul 2013



\bibitem{Silva:2009km}
  F.~P.~Silva and K.~Koyama,
  %``Self-Accelerating Universe in Galileon Cosmology,''
  Phys.\ Rev.\ D {\bf 80}, 121301 (2009)  [arXiv:0909.4538 [astro-ph.CO]].
  %%CITATION = ARXIV:0909.4538;%%  %93 citations counted in INSPIRE as of 11 Jul 2013

%\cite{DeFelice:2010pv}
\bibitem{DeFelice:2010pv}
  A.~De Felice and S.~Tsujikawa,
  %``Cosmology of a covariant Galileon field,''
   Phys.\ Rev.\ Lett.\  {\bf 105}, 111301 (2010)  [arXiv:1007.2700 [astro-ph.CO]].
    %%CITATION = ARXIV:1007.2700;%%  %84 citations counted in INSPIRE as of 11 Jul 2013

%\cite{Gannouji:2010au}
\bibitem{Gannouji:2010au}
  R.~Gannouji and M.~Sami,
  %``Galileon gravity and its relevance to late time cosmic acceleration,''
   Phys.\ Rev.\ D {\bf 82}, 024011 (2010)  [arXiv:1004.2808 [gr-qc]].
    %%CITATION = ARXIV:1004.2808;%%  %62 citations counted in INSPIRE as of 11 Jul 2013


%\cite{DeFelice:2011bh}
\bibitem{DeFelice:2011bh}
  A.~De Felice and S.~Tsujikawa,
  %``Conditions for the cosmological viability of the most general scalar-tensor theories and their applications to extended Galileon dark energy models,''
   JCAP {\bf 1202}, 007 (2012)  [arXiv:1110.3878 [gr-qc]].
   %%CITATION = ARXIV:1110.3878;%%  %27 citations counted in INSPIRE as of 11 Jul 2013

%\cite{Leon:2012mt}
\bibitem{Leon:2012mt}
  G.~Leon and E.~N.~Saridakis,
  %``Dynamical analysis of generalized Galileon cosmology,''
  JCAP {\bf 1303}, 025 (2013)  [arXiv:1211.3088 [astro-ph.CO]].
   %%CITATION = ARXIV:1211.3088;%%  %7 citations counted in INSPIRE as of 11 Jul 2013


%\cite{Adak:2013vwa}
\bibitem{Adak:2013vwa}
  D.~Adak, A.~Ali and D.~Majumdar,
  %``Late time acceleration in a slow moving galileon field,''
  Phys.\  Rev.\  D 88, {\bf 024007} (2013)  [arXiv:1305.2330 [astro-ph.CO]].
   %%CITATION = ARXIV:1305.2330;%%


%\cite{Appleby:2011aa}
\bibitem{Appleby:2011aa}
  S.~Appleby and E.~V.~Linder,
  %``The Paths of Gravity in Galileon Cosmology,''
  JCAP {\bf 1203}, 043 (2012)  [arXiv:1112.1981 [astro-ph.CO]].
   %%CITATION = ARXIV:1112.1981;%%  %18 citations counted in INSPIRE as of 11 Jul 2013

%\cite{Dvali:2000hr}
\bibitem{Dvali:2000hr}
  G.~R.~Dvali, G.~Gabadadze and M.~Porrati,
  %``4-D gravity on a brane in 5-D Minkowski space,''
   Phys.\ Lett.\ B {\bf 485}, 208 (2000)  [hep-th/0005016].
   %%CITATION = HEP-TH/0005016;%%  %1743 citations counted in INSPIRE as of 11 Jul 2013




%\cite{Hossain:2012qm}
\bibitem{Hossain:2012qm}
  M.~.W.~Hossain and A.~A.~Sen,
  %``Do Observations Favour Galileon Over Quintessence?,''
   Phys.\ Lett.\ B {\bf 713}, 140 (2012)  [arXiv:1201.6192 [astro-ph.CO]].
   %%CITATION = ARXIV:1201.6192;%%  %4 citations counted in INSPIRE as of 11 Jul 2013

%\cite{Bartolo:2013ws}
\bibitem{Bartolo:2013ws}
  N.~Bartolo, E.~Bellini, D.~Bertacca and S.~Matarrese,
  %``Matter bispectrum in cubic Galileon cosmologies,''
  JCAP {\bf 1303}, 034 (2013)  [arXiv:1301.4831 [astro-ph.CO]].
   %%CITATION = ARXIV:1301.4831;%%  %4 citations counted in INSPIRE as of 15 Jul 2013



%\cite{Bellini:2013hea}
\bibitem{Bellini:2013hea}
  E.~Bellini and R.~Jimenez,
  %``The parameter space of Cubic Galileon models for cosmic acceleration,''
  arXiv:1306.1262 [astro-ph.CO].
  %%CITATION = ARXIV:1306.1262;%%  %1 citations counted in INSPIRE as of 12 Jul 2013

%\cite{Nesseris:2010pc}
\bibitem{Nesseris:2010pc}
  S.~Nesseris, A.~De Felice and S.~Tsujikawa,
  %``Observational constraints on Galileon cosmology,''
  Phys.\ Rev.\ D {\bf 82}, 124054 (2010)  [arXiv:1010.0407 [astro-ph.CO]].
   %%CITATION = ARXIV:1010.0407;%%  %58 citations counted in INSPIRE as of 12 Jul 2013


%\cite{Jamil:2013yc}
\bibitem{Jamil:2013yc}
  M.~Jamil, D.~Momeni and R.~Myrzakulov,
  %``Observational constraints on non-minimally coupled Galileon model,''
   Eur.\ Phys.\ J.\ C {\bf 73}, 2347 (2013)  [arXiv:1302.0129 [physics.gen-ph]].
    %%CITATION = ARXIV:1302.0129;%%  %1 citations counted in INSPIRE as of 11 Jul 2013


%\cite{Lee:2010tm}
\bibitem{Lee:2010tm}
  H.~W.~Lee, K.~Y.~Kim and Y.~S.~Myung,
  %``Equations of State in the Brans-Dicke cosmology,''
  Eur.\ Phys.\ J.\ C {\bf 71}, 1585 (2011)  [arXiv:1010.5556 [hep-th]].
   %%CITATION = ARXIV:1010.5556;%%  %8 citations counted in INSPIRE as of 11 Jul 2013

%\cite{Lee:2011tb}
\bibitem{Lee:2011tb}
  H.~W.~Lee, K.~Y.~Kim and Y.~S.~Myung,
  %``Future cosmological evolution in $f(R)$ gravity using two equations of state paramters,''
   Eur.\ Phys.\ J.\ C {\bf 71}, 1748 (2011)  [arXiv:1106.2865 [hep-th]].
    %%CITATION = ARXIV:1106.2865;%%  %4 citations counted in INSPIRE as of 12 Jul 2013

\end{thebibliography}
\end{document}